\documentclass[preprints,article,accept,moreauthors,pdftex,10pt,a4paper]{Definitions/mdpi} 

\firstpage{1} 
\pubvolume{xx}
\issuenum{1}
\articlenumber{5}
\pubyear{2022}
\copyrightyear{2022}
\history{}

\Title{Particle creation and the Schwinger model}

\Author{Jos\'e Navarro-Salas$^{1}$
, and Silvia Pla$^{2}$}

\AuthorNames{Firstname Lastname, Firstname Lastname and Firstname Lastname}

\AuthorNames{Firstname Lastname, Firstname Lastname and Firstname Lastname}

\address{%
$^{1}$ \quad Departamento de F\'isica Te\'orica and IFIC, Centro Mixto Universidad de Valencia-CSIC, 
Facultad de F\'isica, Universidad de Valencia, Burjassot-46100, Valencia, Spain; jnavarro@ific.uv.es\\
$^{2}$ \quad Department of Physics, King's College London, Strand Building, Strand Campus, Strand, London.  WC2R 2LS United Kingdom; silvia.pla\_garcia@kcl.ac.uk}

\abstract{We study the particle creation process in the Schwinger model coupled with an external classical source. One can approach the problem by taking advantage that the full quantized model is solvable and equivalent to a (massive) gauge field with a non-local effective action. Alternatively, one can also face the problem following the standard semiclassical route. This means quantizing the massless Dirac field and considering the electromagnetic field as a classical background.  
We evaluate the energy created by a generic,  homogeneous, and time-dependent source. The results exactly match in both approaches. This proves in a very direct and economical way the validity of the semiclassical approach for the (massless) Schwinger model, in agreement with previous analysis based on the linear response equation. Our discussion suggests that a similar analysis for the massive Schwinger model could be used as a non-trivial laboratory to confront a fully quantized solvable model with its semiclassical approximation, therefore mimicking the long-standing confrontation of quantum gravity with quantum field theory in curved spacetime.  } 


\newcommand{\be}{\begin{equation}}
\newcommand{\ee}{\end{equation}}
\newcommand{\bea}{\begin{eqnarray}}
\newcommand{\eea}{\end{eqnarray}}
\usepackage{color}
\newcommand{\ben}{\begin{enumerate}}
\newcommand{\een}{\end{enumerate}}
\usepackage{amsfonts}
\usepackage{amssymb}
\usepackage{amsmath}

\begin{document}


\section{Introduction}

An understanding of the validity domain of the semiclassical approximation has always been an issue in any approach to quantum gravity, especially by those inspired by the theory of quantized fields in curved spacetimes \cite{Hu-Verdaguer, parker-toms}. In this context, it is very useful to analyze toy models for which we have the exact solution to the fully quantized theory, and, at the same time, we can also apply the semiclassical description of the theory, in which we quantize the matter fields but keep a classical description for the background field. The direct comparison between the exact quantum solution and the semiclassical picture offers a useful test for the latter. In this work, we want to carry out this comparison for the Schwinger model of two-dimensional electrodynamics, namely a massless Dirac field coupled to an electromagnetic field  in $1+1$ dimensions \cite{Schwingermass}
\be \label{Schwinger}
S = \int d^2x[-\frac{1}{4}F_{\mu \nu}F^{\mu \nu}  +i\Bar{\psi}\gamma^{\mu}(\partial_{\mu}+ ie A_\mu) \psi +  A_\mu J^\mu]\, .
\ee
For convenience we have also added a classical external source $J^\mu$ coupled to the electromagnetic field $A_\mu$.\\

The semiclassical theory for the massive Schwinger model has been recently analyzed in \cite{Validity}, and the validity of the semiclassical approximation has been critically studied on the basis of the linear response equation. The analysis strongly suggests that the semiclassical theory is a good description of the model as the mass of the fermion goes to zero with respect to the natural mass scale of the problem. Therefore, we find it useful to directly prove the validity of the semiclassical approach in the massless limit by direct comparison between the fully quantized version of the (massless) Schwinger model and the semiclassical description. As usual, the semiclassical picture consists in quantizing the matter field $\psi$, obeying the corresponding Dirac equation in an arbitrary background $A_\mu$. A natural question here is the computation of the particle creation induced by the (time-dependent) background fields ($A_\mu$ and $J^\mu$). It is also of major relevance the computation of the renormalized current $\langle \bar \psi \gamma^\mu \psi \rangle$ as well as the renormalized stress-energy tensor $\langle T^{\mu\nu} \rangle$ for the Dirac field in the appropriated vacuum state. One can use  $\langle \bar \psi \gamma^\mu \psi \rangle$ to write down the semiclassical Maxwell equations and solve for $A_\mu$. This provides a  set of coupled equations.
(i.e., Dirac and semiclassical Maxwell equations). 
By solving them, we can predict, for instance, the energy created by the external source $J^\mu$, in the form of quantized matter-energy and classical electromagnetic energy. The crucial question is: what does this have to do with the exact description of the fully quantized Schwinger model? The main aim of this work is to answer this simple question. For simplicity, we will assume a homogeneous external source $J^\mu = J^\mu (t)$. We find that the total energy created by a  generic (homogeneous) and asymptotically trivial source in the semiclassical picture exactly matches the analogous question posed in the fully quantized theory. This strongly supports the idea that the semiclassical approach is a good description, at least for tiny fermionic masses, as suggested in \cite{Validity}. 

\section{Scalar field coupled linearly to an external source}\label{section1}
In this preliminary section we will briefly review, for pedagogical purposes, the elementary model 
\be 
\mathcal{L}= \frac{1}{2}\partial_\mu \phi \partial^\mu \phi - \frac{1}{2}m^2 \phi^2 + J \phi
\ . \ee
The quantized scalar field $\phi$ interacts with some external $c$-number source $J(x)$. This theory, defined in $1+3$ dimensions, has been used as an illustrative solvable model in quantum field theory courses \cite{ColemanLectures,  Preskill} (for older approaches, see \cite{Wentzel, VanHove}). Time-dependent scattering theory can be immediately applied to this problem if the source $J(x)$ is assumed to turn off at early and late times. A further realistic condition is to assume that the source also goes to zero at spatial infinity. The vacuum persistence amplitude \cite{schwinger},  which  can be expressed in terms of the scattering matrix $S$ as
\be Z[J]= \langle 0 | S | 0 \rangle = e^{iW[J]} \ , \ee
can  be worked out exactly using the interaction picture and Wick's theorem. 

 In the interaction picture the field $\phi_I$ evolves as  $ (\Box + m^2)\phi_I = 0$.  For simplicity we will omit the index $I$ in $\phi_I(x)$ and we will treat $\phi_I\equiv \phi$ as a free field. The Wick algorithm implies
\bea S= I + i\int d^4x_1 :\phi(x_1): J(x_1) -\frac{1}{2!} \int d^4x_1d^4x_2 : \phi(x_1)\phi(x_2): J(x_1)J(x_2) \\ \nonumber
-\frac{1}{2!} \int d^4x_1d^4x_2 D_F(x_1-x_2) J(x_1)J(x_2) + \mathcal{O}(J^3) \ . \eea
$D_F(x_1-x_2)$ is the Feynman propagator of the scalar field.
We can arrange all terms in the form
\be {S}=\,: e^{i\int d^4x \phi(x)J(x)}: \ e^{-\frac{1}{2} \int d^4x_1d^4x_2 D_F(x_1-x_2) J(x_1) J(x_2)} \ . \ee

To get a more explicit expression it is convenient to introduce the Fourier transform of $J(x)\equiv J(t, {\bf x})$, 
where $\tilde J(\omega_{p}, {\bf p})= \tilde J(\omega=\sqrt{m^2 + \bf p^2}, {\bf p})$, and
\be \tilde J(\omega_{p}, {\bf p}) = \int d^4x\,  e^{i(\omega_p t - \bf p \cdot \bf x)} J(t, {\bf x}) \ , \ee 
and write the free-field operator $\phi$ as the usual Fourier expansion
\be  \phi (x)= \int \frac{d^3p}{(2\pi)^3 \sqrt{2\omega_p}} (a_{\bf p} e^{-ipx} + a_{\bf p}^{\dagger} e^{ipx}) \ . \ee
Therefore
\be {S}= {e^{i \int \frac{d^3p}{(2\pi)^3 \sqrt{2\omega_p}} a_{\bf p}^\dagger \tilde J(\omega_p, {\bf p})}   e^{i \int \frac{d^3p}{(2\pi)^3 \sqrt{2\omega_p}} a_{\bf p}\tilde J^*(\omega_p, {\bf p})}} \ e^{-\frac{1}{2} \int d^4x_1d^4x_2 D_F(x_1-x_2) J(x_1) J(x_2)} \ . \ee

From the above expressions it is not difficult to get
\be W[J]= \frac{i}{2}\int d^4x_1 d^4x_2 J(x_1)D_F(x_1-x_2)J(x_2)  \ . \ee It is found that the effective action $W[J]$ has, in general, a non-zero imaginary part
\be W[J]= i\frac{\alpha}{2} + \frac{\beta}{2}
\ , \ee
where
\be  \operatorname{Re} \Bigg[ \int d^4x_1d^4x_2 D_F(x_1-x_2) J(x_1) J(x_2)\Bigg] \equiv \alpha \ . \ee
The coefficient $\beta$  appears as a pure phase in the $S$ matrix and, in our context, it has no relevant physical meaning. The model is exactly solvable and the probability that $n$ particles are created by the source follows a Poisson distribution, with mean value
\be \langle N \rangle = \alpha = \int \frac{d^3{\bf p}}{2(2\pi)^3\omega_{p}} |\tilde J(\omega_{p}, {\bf p})|^2 \ . \ee
If the source is parametrized by a time-dependent function $f(t)$ times a ``static source'' $j(\bf x)$, i.e., $J(x) = f(t) j(\bf x)$,  the $S$-matrix tends to the identity in the adiabatic limit ($f(t) \to 1$). In other words, the number of created particles $\langle N \rangle$ vanishes for a purely static source, in accordance  with  energy conservation. 

\section{Schwinger model coupled to an external source}

The Schwinger model describes the interaction of massless charged fermions  in $1+1$ dimensions.  It is defined by the action
\be \label{Schwinger}
S = \int d^2x[-\frac{1}{4}F_{\mu \nu}F^{\mu \nu}  +i\Bar{\psi}\gamma^{\mu}(\partial_{\mu}+ ie A_\mu) \psi +  A_\mu J^\mu]\quad ,
\ee
where $e$ is the electric charge of the fermion. In two dimensions, the electric charge has dimension of mass. For future purposes, we have also added an external and conserved classical source $J^\mu$,  coupled to the electromagnetic field. The theory is invariant under local $U(1)$ gauge transformations as well as a global chiral transformation. 

The quantized model can be solved exactly \cite{Schwingermass} and it turns out to be equivalent to a  pseudoscalar field with a non-zero mass given by $m_\gamma^2=e^2/\pi$. Using functional integral methods \cite{Das}, the fully quantized theory can also be displayed in terms of a free massive vector field with effective action  
\be \label{proca} S_{eff}= \int d^2x [-\frac{1}{4} F_{\mu\nu}F^{\mu\nu} + \frac{1}{2}\frac{e^2}{\pi}A_\mu(\eta^{\mu\nu} -\partial^\mu \Box^{-1} \partial^\nu) A_\nu] + A_\mu J^\mu]\ , \ee
where $S_{eff}$ has been defined as
\be
    \label{eff}e^{i S_{eff}[A_\mu; J_\nu]} =  \int D\bar \psi D\psi \ e^{i S[A_\mu, \bar \psi, \psi; J_\nu]} \quad .
\ee

In the Lorenz gauge, the effective field equations are equivalent to the equations for a massive vector field (i.e., a Proca field)
\be \partial_\mu F^{\mu\nu} + \frac{e^2}{\pi}A^\nu= J^\nu \ . \ee 
The Coulomb force is then replaced by a force of range determined by the mass scale $m_\gamma=|e|/\sqrt{\pi}$.

As in the model described in Section \ref{section1}, the vacuum persistence amplitude is given by 
\be 
\langle 0| S | 0 \rangle \equiv Z[J] = \ e^{-\frac{1}{2} \int d^2x_1d^2x_2 D_{\mu\nu}(x_1-x_2) J^\mu(x_1) J^\nu(x_2)}   \ , \ee
where $D_{\mu\nu}(x_1-x_2)$ is the covariant Feynman propagator for the Proca field

\be  D_F^{\mu \nu}(x-y)  = i [\eta^{\mu\nu} (\Box+ m_\gamma^2)-\partial^\mu\partial^\nu]^{-1} \delta^{(2)}(x-y)= \int \frac{d^2p}{(2\pi)^2} \frac{i(\frac{p_\mu p_\nu}{m_\gamma^2} -\eta_{\mu\nu})}{p^2  -m_\gamma^2 +i\epsilon} e^{-ip(x-y)} \ . \ee
As it is well-known, it can be re-expressed as 
\be D_F^{\mu \nu}(x-y)  = (-\eta^{\mu\nu} + m_\gamma^{-2} \partial^\mu \partial^\nu) D_F(x-y)\, , \ee
where $D_F(x-y)$ is the Feynman propagator of a scalar field of mass $m$ (in our case $m_\gamma=|e|/\sqrt{\pi}$). Since the Proca propagator is coupled to external conserved currents we can ignore the term with total derivatives, so

\be 
\langle 0| S | 0 \rangle \equiv Z[J] = \ e^{-\frac{1}{2} \int d^2x_1d^2x_2 D_F(x_1-x_2) (-\eta_{\mu\nu} J^\mu(x_1) J^\nu(x_2))}    . \ee

Defining the Fourier components of the current (for now on, in our $1+1$ dimensional model ${\bf p} \equiv p $, $\omega_p =\sqrt{m^2 + p^2}$, and ${\bf x} \equiv x$)
\be   \int d^2x\,  e^{{i(\omega_p t - px)} } J^\mu(t,x)  \equiv \tilde J^\mu(\omega_p, {\bf p}) \equiv \tilde J^\mu (p) \ , \ee
it can be shown that the spectrum of  particles created by the source at late-times is also a Poisson distribution. The probability of creating $n$ bosonic particles is  
\be P(n) = e^{-\alpha} \frac{\alpha^n}{n!} \ , \ee
where $e^{-\alpha} = P(0)$ and 
\be \alpha =  \int_{-\infty}^{+\infty}\frac{d p}{(2\pi) 2\omega_{p}}  |\tilde J^\mu ({p})\tilde J_\mu^* ({ p})|.\ee
Therefore, the probability that vacuum remains the vacuum is 
 $ | \langle 0| S | 0 \rangle |^2 = P(n=0) = e^{-\alpha}$ and the mean value of the number of created particles is given by
\be \langle N \rangle \equiv \sum_{n=0}^\infty n P(n)= \alpha. \ee
 Furthermore,  using similar results as those presented in the previous section it is not difficult to show that the mean value of the energy of the created particles is ($H$ is the Hamiltonian of the theory in the interacting picture)
\be \langle 0| S^\dagger H S |0\rangle =  \int_{-\infty}^{+\infty}\frac{d{p}}{(2\pi) 2\omega_{p}} \omega_p |\tilde J^\mu ({ p})\tilde J^*_\mu ({p})|.\ee

\subsection{Production of Proca particles}
With these general results, we can directly analyze the particle production phenomena induced by some specific configurations. 
\subsubsection{Delta profile}
Let us start with a very simple case. The case where the external current is homogeneous and its time dependence is given by the dirac delta, i.e. $J^\mu(t,x) = (0, -E_0 \delta (t))$. The Fourier transform of the classical current is given by
\be   \int d^2x \, e^{i(\omega_p t - px)} J^\mu(t,x)  
= (0, -2\pi E_0 \delta({p}))\ . \ee
Therefore, 
\be \alpha =  \int_{-\infty}^{+\infty}\frac{d{p}}{(2\pi) 2\omega_{p}} (2\pi E_0)^2 (\delta({p}))^2 = \frac{2\pi E_0^2}{2\omega_{0}} \delta(0)\ . \ee
Taking into account that $\omega_{0}\equiv \omega_{p=0}$ is the mass of the Proca field $m_\gamma= \frac{|e|}{\sqrt{\pi}}$ and that we can relate the delta function at $p=0$ with the length $L$ of one-dimensional space
$ \delta(0) = L/2\pi$
we get
\be \alpha = \frac{E_0^2 L}{2\omega_{0}}= \frac{E_0^2 L}{2|e|/\sqrt{\pi}} \ . \ee
The (late time) number density is then given by
\be \langle n \rangle= \frac{E_0^2}{2|e|/\sqrt{\pi}} \ . \ee
In the same way, the energy of the created particles is 
\be  \langle 0| S^\dagger H S |0\rangle =  \int_{-\infty}^{+\infty}\frac{d{p}}{(2\pi) 2\omega_{p}} \omega_p |\tilde J^\mu ({ p})\tilde J^*_\mu ({p})| = \frac{E_0^2}{2}L \ , \ee
and hence the corresponding energy density of the created particles at late time is 
\be  \langle 0| S^\dagger T_{00}  S |0\rangle = \frac{E_0^2}{2}\ . \ee

\subsubsection{Spatially homogeneous  profile $J^\mu(t)$}
We can evaluate the 
energy density for an arbitrary spatially homogeneous profile $J^\mu(t)= (0, J^1(t))$. First, we have to determine the current in momentum space
\be   \int d^2x e^{i(\omega_p t - px)} J^\mu(t,x)  
= (0, 2\pi \delta({p})) \int_{-\infty}^{+\infty} dt J^1(t) e^{i\omega_{0}t} \ . \ee

Therefore, the energy of the created particles is
\bea  \langle 0| S^\dagger H S |0\rangle &=&  \int_{-\infty}^{+\infty}\frac{d{p}}{(2\pi)^2} |\tilde J^\mu ({ p})\tilde J^*_\mu ({p})| \nonumber \\
&=& \frac{1}{4\pi} \int_{-\infty}^{+\infty} d{p} (2\pi)^2 \delta({p}) \frac{L}{2\pi}\Bigg| \int_{-\infty}^{+\infty} dt J^1(t) e^{i\omega_{0}t}\Bigg|^2 \nonumber \\
&=& \frac{L}{2} \Bigg| \int_{-\infty}^{+\infty} dt J^1(t) e^{i\omega_{0}t}\Bigg|^2\ . \eea
And the energy density of the produced Proca particles reads  ($T_{\mu\nu}$ is the stress-energy tensor) 
\be \label{eq:final-En-Schwinger} \langle 0| S^\dagger T_{00} S |0\rangle= \frac{1}{2} \Bigg| \int_{-\infty}^{+\infty} dt J^1(t) e^{i\omega_{0}t}\Bigg|^2 \ . \ee
The above general result can be worked out with more detail for specific forms of the time-dependent profile $J^1(t)$.
For example, for the Sauter current $J^1 (t)=2eE_0 \operatorname{sech}^2(\sigma t)\tanh(\sigma t)$, the energy density of the produced Proca particles is \be \langle 0| S^\dagger T_{00} S |0\rangle= \frac{e^6E_0^2 }{2\sigma^6}\operatorname{cosech}\Big(\frac{|e|\sqrt{\pi}}{2\sigma}\Big) \ . \ee

\subsubsection{ Time-independent profiles $J^\mu (x)$}

There is no particle creation for these profiles.

\section{Semiclassical approximation}

We can now proceed to analyze the problem of particle creation induced by time-varying electric fields through the semiclassical approximation. To this end, let us first examine the massive theory.  We recall that in this approach, the matter degrees of freedom are quantized using canonical quantization. As usual in this context, we work in the Heisenberg picture. The electromagnetic field is treated as a classical background.  

Consider a quantized spin-$\frac{1}{2}$ field $\psi$ in a two-dimensional Minkowski spacetime, coupled with a classical, homogeneous electric field so that $E=E(t)$ in a given reference frame. It can be described in terms of the vector potential  $A_\mu=(0,-A(t))$ in the Lorenz gauge. 
The Dirac equation for the  field reads
\bea
(i\gamma^{\mu}D_\mu-m)\psi=0\, ,
\eea
where $D_\mu \psi=(\partial_\mu+ieA_\mu)\psi$ 
and where $\gamma^\mu$ are the gamma matrices which satisfy the anticomutation relations $\{\gamma^\mu,\gamma^\nu\}=2\eta^{\mu \nu}$. From now on, we use the Weyl representation for the gamma matrices. Since the potential vector is homogeneous, one can expand the fermionic field in modes as $\psi=\int_{-\infty}^\infty d{ p}[B_{p}u_p(t,x)+D^{\dagger}v_{ p}(t,x)]$, where the two independent spinor solutions can be written as
\bea
 u_{{p}}(t, x)=\frac{e^{ipx}}{\sqrt{2\pi }} \scriptsize \left( {\begin{array}{c}
 h^{I}_p(t)   \\
 -h^{II}_p (t) \\
 \end{array} }\right)\ , \qquad 
  v_{p}(t, x)=\frac{e^{-ipx}}{\sqrt{2\pi }} \scriptsize \left( {\begin{array}{c}
 h^{II*}_{-p} (t)  \\
 h^{I*}_{-p}(t)  \\
 \end{array} } \right)
 \label{spinorde}
\ , \eea
and where $B_p$, $B^\dagger_p$, $D_p$ and $D^\dagger_p$ are the creation and annihilation operators, which fulfill the usual anticomutation relations. Inserting the mode expansion into the Dirac equation, we directly find
\bea
\dot{h}^{I}_p-i\left(p-eA\right)h^{I}_p-i m h^{II}_p=0\label{mod1}\\ 
\dot{h}^{II}_p+i\left(p-eA\right)h^{II}_p-i m h^{I}_p=0 \label{mod2}
\eea
together with the normalization condition $|h_p^I|^2+|h_p^{II}|^2=1$.

On the other hand, the semiclassical Maxwell equations for $F_{\mu\nu}=\partial_\mu A_\nu-\partial_\nu A_\mu$ are 
\bea \label{eq:semimax0}
\partial_\mu F^{\mu \nu}=J^\nu+\langle J^\nu_Q\rangle_{\rm ren}.
\eea
Here, we have directly split the source term in two parts: $J^\nu$ is a prescribed (and conserved) classical source of the form $J^{\nu}=(0,J^1(t))$, and $\langle J^\nu_Q\rangle$ is the induced Dirac current 
$J^\nu_Q=e\bar \psi\gamma^\nu \psi$. The vacuum expectation value of this observable is ultraviolet divergent and has to be regularized in a proper way. Here, since we work with an homogeneous background, it is very convenient to use the adiabatic renormalization method, originally proposed for scalar fields, but it can be consistently extended for two-dimensional fermions interacting with electromagnetic backgrounds. The complete procedure is explained in \cite{FN,FNP} (for a comparison among different renormalization methods, see \cite{Silvia}). 
After subtracting the appropriate terms, one finds 
\bea
\langle J^0_Q \rangle_{\rm ren}&=&0\ ,\\
\langle J^1_Q \rangle_{\rm ren}&=&-\frac{e}{2\pi}\int_{-\infty}^{\infty}dp\Big(|h_p^{II}|^2-|h_p^{I}|^2-\frac{p}{w}\Big) -\frac{e^2}{\pi}A\ ,\label{currentRen}\eea
where we recall that $w=\sqrt{p^2+m^2}$ and $p\equiv p^1$. The expression for $\mu=0$ corresponds to the induced electric charge and, as expected, it is identically zero: no net charge is created.  On the other hand, the expression of the electric current ($\mu=1$)  reflects the existence of created particles induced by the external background. 
Taking all these results into account, the only non-vanishing Maxwell equation \eqref{eq:semimax0} is 
\bea
\ddot A+\frac{e^2}{\pi}A=J^1(t)-\frac{e}{2\pi}\int_{-\infty}^{\infty}dp\Big(|h_p^{II}|^2-|h_p^{I}|^2-\frac{p}{w}\Big)\,.\label{Max1}
\eea
Equations \eqref{Max1}, \eqref{mod1} and \eqref{mod2}, together with the normalization condition are the so called {\it semiclassical backreaction equations}. We can also compute the vacuum expectation value of the energy density of the Dirac field \cite{FNP}
\bea \label{T00ren}
\langle \rho \rangle _{\rm ren}=\frac{1}{2\pi} \int_{-\infty}^{\infty} d p\Bigg[ i\left(h_{p}^{I I} \dot{h}_{p}^{I I *}+h_{p}^{I} \dot{h}_{p}^{I *}\right)+w-\frac{eA  p}{w}\Bigg]+\frac{e^2A^2}{2\pi}\, .
\eea

Now, we can proceed to study the massless limit. In this case the mode equations \eqref{mod1} and \eqref{mod2} decouple, and their solutions are given by $h^{I,II}_p(t)=\pm \theta(\mp p)e^{\pm i\int_{t_0}^t(p-eA(t'))dt'}$, where $\theta(x)$ is the Heaviside step function. Moreover, the induced electric current reduces to $\langle J^1_Q\rangle_{\rm ren}=-\frac{e^2}{\pi}A $. Hence, the semiclassical Maxwell equation (\ref{Max1}) turns out to be 
\bea \label{semi-maxwell}
\ddot A+\frac{e^2}{\pi}A=J^1(t)\, ,
\eea
that is, the equation of a harmonic oscillator with frequency $\omega_0=\frac{|e|}{\sqrt{\pi}}$ coupled to an external source $J^1(t)$. We note that  this result is consistent with the fact that radiative corrections in two-dimensional electrodynamics induce a mass for the photon \cite{Schwingermass, Peskin-Schroeder}. 
The energy density of the fermion field (\ref{T00ren}) can also be  reduced to
\be\langle \rho \rangle_{\rm ren}=\frac{e^2}{2\pi}A^2 \ .  \label{enN}\ee

\subsection{Late times energy density for an asymptotically bounded profile}
In order to compare the semiclassical model with the fully quantized Schwinger model, it is very useful to compute the total energy density in the semiclassical framework induced by an asymptotically bounded external source $J^1(t\to \pm \infty)=0$. 
At late times, the energy density of the semiclassical system $\mathcal{E}=\langle \rho \rangle_{\textrm{ren}} +\rho_{\textrm{elec}}$ should be a conserved quantity, namely 
\be \label{eq:out-En}
\mathcal{E}_{\textrm{out}}=\frac{\omega_0^2}{2 }A(t)^2_{\textrm{out}}+\frac{1}{2}E(t)^2_{\textrm{out}}=cte\, .
\ee
It is not difficult to compute the explicit value of $\mathcal{E}_{\textrm{out}}$. The general solution of \eqref{semi-maxwell}, satisfying the boundary conditions at early times ($t_0\to - \infty$)  $A(t_0)=\dot A(t_0)=0$,  is given by
\bea \label{eq:Ageneral}
A(t)= - \frac{\cos(\omega_0 t)}{\omega_0}\int^t_{-\infty} J^1(t')\sin(\omega_0 t') dt' + \frac{\sin(\omega_0 t)}{\omega_0}\int_{-\infty}^t J^1(t')\cos(\omega_0 t') dt'.
\eea
Therefore, the late-times asymptotic behavior of the vector potential \eqref{eq:Ageneral} reads
\be
A(t)_{\textrm{out}}=\frac{1}{\omega_0}\int_{-\infty}^{\infty}\sin\big(\omega_0(t-t')\big)J^1(t')dt'=\frac{1}{\omega_0}\operatorname{Im}\Bigg[e^{i\omega_0 t}  \int_{-\infty}^{\infty} e^{-i \omega_0 t'}J^1(t')dt'\Bigg]\ .
\ee
From this, we easily get the asymptotic behavior of the electric field is 
\bea \label{E-inf-2}
E(t)_{\textrm{out}}= -\int^\infty_{-\infty}\cos\big(\omega_0 (t-t')\big)J^1(t') dt'= -\operatorname{Re}\Bigg[e^{i\omega_0 t}  \int_{-\infty}^{\infty} e^{-i \omega_0 t'}J^1(t')dt'\Bigg]\ .
\eea
Consequently, the total energy density at late-times becomes \eqref{eq:out-En}
\be\label{eq:Eout}
\mathcal{E}_{\textrm{out}}=\frac{1}{2} \Bigg|  \int_{-\infty}^{\infty} e^{-i \omega_0 t'}J^1(t')dt'\Bigg|^2\, .
\ee
This result is in complete agreement with the one obtained in the fully quantized theory \eqref{eq:final-En-Schwinger}.

\section{Conclusions and final comments}

One of the major problems in theoretical physics is reconciling gravity with the quantum theory. An intermediate step in this direction is provided by the theory of quantized fields in curved spacetime and semiclassical gravity, where the spacetime metric is treated classically. This approach, while incomplete, has provided many insights into the understanding of physical processes in the early universe and black holes, most of them related to the creation of particles or perturbations \cite{parker-toms, Hu-Verdaguer}. In the absence of a complete and self-consistent quantum theory of gravity, the confrontation of it (the ``right answer'') with the semiclassical approach has always been considered provisional.  An example of this tension is provided by the information loss problem of evaporating black holes, where a wide spectrum of viewpoints has emerged since the mid-seventies \cite{Unruh-Wald}. 
In general, the validity of semiclassical gravity has been analyzed from different perspectives, and attempts to address this question from the semiclassical description itself are likely the best strategy (see \cite{Anderson1} and references therein).

Another viewpoint to analyze the validity of the semiclassical approximation of a  theory is by direct comparison between the exact solution of the fully quantized theory (when it is available)  with its semiclassical description. We have substantiated this viewpoint by studying a toy model, namely the two-dimensional (massless) Schwinger model. We have analyzed, within this solvable model, the issue of particle production by a time-dependent, homogeneous source. We have checked explicitly that the semiclassical method provides the exact, ``right answer'' when it is evaluated in terms of the energy produced by the external source. 
 The time-dependent external source $J^1(t)$ excites the vacuum and creates quanta. This phenomenon can be described in two ways. Since the exact theory is equivalent to a massive vector field (or, in $1+1$ dimensions, to a  massive scalar field) the most interesting piece of information is the total produced energy. It is given by Eq. (\ref{eq:final-En-Schwinger}). On the other hand, the semiclassical picture offers an indirect way to re-evaluate this quantity. We can compute the energy created as  fermionic quanta. After renormalization, it is given by  
$\langle \rho \rangle_{\rm ren}=\frac{e^2}{2\pi}A_{out}^2$. But this is not the total amount of created energy, we have to add the contribution of the classical electromagnetic field, namely $ \frac{1}{2} E^2_{out}$, otherwise the computation will miss a very important ingredient.   Only when both quantities are added [see Eq. \eqref{eq:Eout}] we nicely reproduce the prediction (\ref{eq:final-En-Schwinger}) of the fully quantized theory.  We have to stress here that, in this context, it is 
meaningless to compare the number of created Proca particles (in the fully quantized theory) with the number of created fermions living together with a time-dependent electromagnetic field (in the semiclassical theory). 

We leave for a future project to extend our analysis 
to the massive Schwinger model. The model is then much more involved, but it can also be treated analytically in the full quantum picture \cite{Coleman, Coleman2}. On the other hand, the semiclassical Maxwell equations can also be solved in this context using numerical techniques \cite{Kluger-fermions,Validity}, and the validity of this approach can be also studied by means of the linear response equations \cite{Validity}. It was found that the accuracy of the semiclassical approximation decreases as the mass of the Dirac fermions increases. Therefore, the exact agreement between both approaches is expected to fail when the fermions are massive. Nevertheless, for very massive fermions, the semiclassical approximation is again expected to be accurate by the decoupling theorem.  It would be very interesting to directly confront  the semiclassical and the exact approach for the massive Schwinger model, many relevant physics could be expected to be learned. In this brief paper we have paved the way in this direction by analyzing the massless model. This is the underlying motivation of this work.

\authorcontributions{Conceptualization, J. N.-S. and S. P.; investigation, J. N.-S. and S. P.; writing-original draft preparation, J. N.-S. and S. P.; writing-review and editing, J. N.-S. and S. P. All authors have read and agreed to the published version of the manuscript.}

\funding{The work of S.P.~is supported by the Leverhulme Trust, Grant No.~RPG-2021-299. Part of this work is supported by the Spanish Grant  PROMETEO/2020/079 (Generalitat Valenciana), and
 PID2020-116567GB-C2-1 funded by MCIN/AEI/10.13039/501100011033.}
\acknowledgments{ We thank Paul R. Anderson and Ian M. Newsome for useful discussions.}

\conflictsofinterest{The authors declare no conflict of interest.} 


\reftitle{References}





\end{document}